\documentclass{article}
\usepackage{spconfa4,amsmath,graphicx}

\usepackage{comment}

\usepackage{lipsum}

\usepackage{amsfonts}
\usepackage{algorithmic}
\usepackage{algorithm}
\usepackage{array}
\usepackage[caption=false,font=normalsize,labelfont=sf,textfont=sf]{subfig}
\usepackage{textcomp}
\usepackage{stfloats}
\usepackage{url}
\usepackage{verbatim}
\usepackage{graphicx}
\usepackage[sort]{cite}
\usepackage{color}
\usepackage{xcolor}
\usepackage{tabularray}
\usepackage{tikz}
\usetikzlibrary{positioning}
\usepackage{adjustbox}
\usepackage{enumitem}   
\usepackage{pbalance}

\usepackage{booktabs}
\usepackage{hhline}
\usepackage{multirow}
\usepackage{makecell}
\usepackage{diagbox}
\usepackage{svg}
\usepackage{siunitx}
\usepackage[hidelinks]{hyperref} 

\usepackage[nolist]{acronym} 

\begin{acronym}
    \acro{df}[DF]{directivity factor}
    \acro{drr}[DRR]{direct-to-reverberant ratio}
    \acro{ndf}[NDF]{neural directional filtering}
    \acro{pesq}[PESQ]{perceptual evaluation of speech quality}
    \acro{rir}[RIR]{room impulse response}
    \acro{rtf}[RTF]{room transfer function}
    \acro{sdr}[SDR]{signal-to-distortion ratio}
    \acro{stft}[STFT]{short-time Fourier transform}
    \acro{vdm}[VDM]{virtual directional microphone}
    \acro{wpe}[WPE]{weighted prediction error}

	\acro{DNN}[DNN]{deep neural network}
    \acro{SRMR}[SRMR]{speech-to-reverberation modulation energy ratio}
    \acro{MOS}[MOS]{mean opinion score}
    
    \acro{CDR}[CDR]{coherent to diffuse ratio}
    \acro{FBF}[FBF]{fixed beamforming}
    \acro{ITD}[ITD]{interaural time difference}
	\acro{DDF}{joint dereverberation and directional filtering}
	\acro{sdri}[$\Delta$SDR]{improvement in \ac{SDR} over the unprocessed signal}
	\acro{DMA}[DMA]{differential microphone array}
	\acro{DNN}[DNN]{deep neural network}
	\acro{DOA}[DOA]{direction-of-arrival}
	\acrodefplural{DOA}[DOAs]{directions-of-arrival}
	
	\acro{iSTFT}[iSTFT]{inverse short-time Fourier transform}
	\acro{CDMA}[CDMA]{circular \ac{DMA}}

	\acro{LDMA}[LDMA]{linear \ac{DMA}}
        \acro{LS}{least-squares}
	\acro{LSTM}[LSTM]{long short-term memory}
        \acro{BiLSTM}[BiLSTM]{bidirectional LSTM}
        \acro{UniLSTM}[UniLSTM]{unidirectional LSTM}
        \acro{WNG}[WNG]{white noise gain}
	\acro{RIRs}[RIRs]{room impulse responses}
	\acro{RIR}[RIR]{room impulse response}
        \acro{RTF}[RTF]{room transfer function}
        \acro{RTFs}[RTFs]{room transfer functions}        
	\acro{DPIR}[DPIR]{direct-path impulse response}
        \acro{MVDR}[MVDR]{minimum variance distortionless response}
        \acro{LCMV}[LCMV]{linear-constraint minimum-variance}
        \acro{PMWF}[PMWF]{parametric multichannel wiener filter}
        \acro{GSC}[GSC]{Generalized sidelobe canceller}
        \acro{FT-JNF}[FT-JNF]{joint spatial and temporal-spectral non-linear filtering}  
        \acro{JNF}[JNF]{joint non-linear filtering }        
        \acro{SSF}[SSF]{spatially selective
deep non-linear filter } 
	\acro{SDR}[SDR]{signal-to-distortion ratio}
        \acro{noisySDR}[reference microphone]{\ac{SDR} of the unprocessed omnidirectional reference microphone}
    \acrodefplural{noisySDR}[$\textrm{SDRs}^\textrm{omni}$]{\acp{SDR} of the unprocessed omnidirectional microphone}
	\acro{SNR}[SNR]{signal-to-noise ratio}
	\acro{STFT}[STFT]{short-time Fourier transform}
         \acro{MAE}[MAE]{mean absolute error}	
	\acro{TF}[TF]{time-frequency}
	\acro{tsdr}[SA-$\varepsilon$-tSDR]{source-aggregated and regularized thresholded \ac{SDR}}
      \acro{STOI}[STOI]{short term objective intelligibility}
    \acro{PESQ}[PESQ]{perceptual evaluation of speech quality}
	
	\acro{UCA}[UCA]{uniform circular array}
        \acro{NDF}[NDF]{neural directional filtering} 
        \acro{SHONDC}[SHONDC]{steerable high-order neural directional coding}
        \acro{NDSC}[NDSC]{neural directional speech coding}
        \acro{NDC}[NDC]{neural directional coding}
        \acro{WNG}[WNG]{white noise gain}
        \acro{DF}[DF]{directivity factor}
        \acro{DI}[DI]{directivity index}
        \acro{HRTF}[HRTF]{head-related transfer function}
        \acro{ILD}[ILD]{interaural level difference}
        \acro{FiLM}[FiLM]{feature-wise linear modulation}     
        \acro{PESQ}[PESQ]{perceptual evaluation of speech quality}
        \acro{UNDF}[UNDF]{neural directional filtering with user-defined directivity patterns} 
    \acro{VDM}[VDM]{virtual directional microphone}
    \acro{DirAC}[DirAC]{directional audio coding}
    \acro{FOA}[FOA]{first-order ambisonics}
    \acro{HOA}[HOA]{high-order ambisonics}
        \acro{ATF}[ATF]{acousitc transfer function}

\end{acronym}

\title{GAN-based Joint Dereverberation and Directional Filtering}
\name{Weilong Huang, Shrishti Saha Shetu, Emanu{\"e}l A. P. Habets}
\address{International Audio Laboratories Erlangen\textsuperscript{$\ast$}, Am Wolfsmantel 33, 91058 Erlangen, Germany\thanks{\textsuperscript{$\ast$}A joint institution of Fraunhofer IIS and Friedrich-Alexander-Universit{\"a}t Erlangen-N{\"u}rnberg (FAU), Germany. The authors gratefully acknowledge the scientific support and HPC resources provided by the Erlangen National High Performance Computing Center (NHR@FAU) of the Friedrich-Alexander-Universität Erlangen-Nürnberg (FAU). The hardware is funded by the German Research Foundation (DFG).}}
\begin{document}
\ninept
\maketitle
\begin{abstract}
Recently, neural directional filtering (NDF) enables reconstruction of a virtual directional microphone (VDM) with a desired directivity pattern, accurately rendering multi-source scenes by preserving spatial cues. In strongly reverberant environments, spatial cues become perceptually difficult to distinguish, limiting NDF-based spatial sound capture. This paper addresses this limitation with three contributions: First, we propose a neural dereverberation and directional filtering (NDDF) approach to reconstruct dereverberated VDM signals. Second, NDDF is implemented with discriminatively trained and generative adversarial network (GAN)-based models, compared with cascaded dereverberation and directional-filtering baselines. Experimental results indicate that the NDDF consistently surpasses the cascaded baselines. Additionally, the GAN-based NDDF outperforms the discriminative variant when addressing a high-order VDM target. Third, we introduce a method for directivity pattern estimation that relies solely on the input and output signals. This method is suitable for signal-mapping-based spatial filtering, which synthesizes the output signal directly without explicit filtering or masking.
\end{abstract}
\section{Introduction}
\label{sec:intro}
Spatial sound capture aims to preserve the spatial cues of an acoustic scene, enabling listeners to perceive source positions and room characteristics during playback \cite{blauert1997spatial}. In enclosed environments, reverberation introduces delayed reflections that overlap with the direct sound, thereby degrading spatial cues such as \ac{ILD}. This degradation is particularly significant when employing a compact array with a small aperture and few microphones, as conventional fixed beamforming (FBF) applied to such arrays yields limited performance \cite{benesty2018fixed}.

Recently, \ac{NDF} has been proposed as a data-driven alternative for reconstructing a \ac{VDM} with a desired directivity pattern on compact arrays \cite{ndf_iwaenc, ftjnf_steerable, NDF}. By learning the input-output behavior of an ideal directional microphone, NDF can achieve a high-directivity frequency-invariant response, and even supports arbitrary directivity pattern configuration at inference \cite{huang2025neural}. However, for both \ac{FBF} and \ac{NDF}, a higher directivity comes with a narrower mainlobe, which is not always desirable for spatial sound capture. Certain recording techniques require a specific shape for the directivity pattern: for instance, the widely used X-Y stereo technique relies on a pair of first-order cardioid patterns \cite{rumsey2014sound}. A first-order cardioid offers a \ac{DI} of only approximately 4.8~dB \cite{eargle2012microphone}, which is insufficient to suppress reverberant energy in strongly reverberant environments. In such scenarios, the reconstructed \ac{VDM} exhibits substantial reverberation, obscuring spatial cues and degrading the perceptual quality of the captured scene. Whether a method can both flexibly realize various directivity patterns like NDF and maintain effective dereverberation under each of these patterns remains an open question. An intuitive remedy is to apply a dereverberation front-end prior to NDF, but such cascaded pipelines optimize each stage independently and are therefore unlikely to yield an optimal final output. This motivates a unified formulation that jointly addresses dereverberation and directional filtering.

In this paper, we propose neural dereverberation and
directional filtering (NDDF), a joint neural approach that reconstructs a dereverberated VDM signal directly from the array input. Our contributions are threefold. First, we formulate joint dereverberation and directional filtering as a single learning problem and implement NDDF with two training paradigms: a discriminatively trained model and a generative adversarial network (GAN)-based model. We compare them against cascaded dereverberation and directional-filtering baselines. Second, experimental results show that the NDDF consistently outperforms the cascaded baselines, where the GAN-based NDDF outperforms the discriminative variant for a high-order VDM target. Third, since the GAN-based NDDF synthesizes the output signal directly without explicit filtering or masking, conventional directivity analysis is not applicable; we therefore introduce a directivity pattern estimation method that relies solely on the input and output signals, which is broadly applicable to signal-mapping-based spatial filtering approaches.

\section{Problem Formulation}
\label{sec:format}
We consider a scenario in which a compact array with $Q$ omnidirectional microphones captures an acoustic scene comprising $N$ sound sources in a reverberant room. Let $X_{q,n}(f,t)$ denote the $n$-th source signal at the $q$-th microphone in the \ac{STFT} domain, where $f$ and $t$ represent the frequency and frame indices, respectively. The mixture signal at the $q$-th microphone, denoted by $Y_q(f,t)$, is given by
\vspace*{-0.15cm} 
    \begin{equation} \label{eqn:mic_sig}
        Y_q(f,t) = \sum_{n=1}^{N} X_{q,n}(f,t) + V_q(f,t),~q\in\{1,2,\ldots,Q\},
    \end{equation}
where $V_q(f,t)$ denotes spatially uncorrelated sensor noise across the microphones. Additionally, $X_{q,n}(f,t) = H_{q,n}(f) \, S_{n}(f,t)$ \cite{avargel2007multiplicative}, where $S_{n}(f,t)$ is the $n$-th source signal and $H_{q,n}(f)$ models the \ac{RTF} between the $n$-th source and the $q$-th microphone.

The \ac{NDF} task employs a \ac{DNN} to reconstruct a \ac{VDM} signal that captures the acoustic scene according to a specified directivity pattern \cite{ndf_iwaenc, NDF}. The \ac{VDM} position is set at the reference microphone ($q=1$). The directivity pattern, represented by $\Lambda(\theta, \phi)$, defines the directional sensitivity of a beamformer or directional microphone and describes the spatial response to sounds arriving from different directions \cite{elko2000superdirectional, eargle2012microphone}. Consequently, the \ac{VDM} signal $Z_{\mathrm{vdm}}(f,t)$ is defined as
\vspace*{-0.15cm} 
    \begin{equation}\label{eqn:vdm_sig_rvb}
        Z_{\mathrm{vdm}}(f,t)= \sum_{n=1}^{N}  
 H_{\mathrm{vdm}, n}(f; \Lambda) \, S_{n}(f,t),
    \end{equation}
where $H_{\mathrm{vdm}, n}(f;\Lambda) =\sum_{i=1}^{\infty} \Lambda(\theta_i, \phi_i) \,  \rho^{(i)}_{\mathrm{vdm},n}[f ] $ represents the \ac{RTF} between the $n$-th source and the \ac{VDM}. The term $\rho^{(i)}_{\mathrm{vdm},n}[f]$ denotes the transfer function of the $i$-th propagation path from the $n$-th source to the \ac{VDM} within a reverberant environment. Each reflection path is weighted by the directivity gain associated with its incident direction. The angles $\theta_i$ and $\phi_i$ specify the incident direction for the $i$-th propagation path. 

To minimize the impact of late reflections (reverberation) on \ac{VDM}, we propose a neural approach that reconstructs a dereverberated \ac{VDM} signal. Specifically, we decompose $H_{\mathrm{vdm}, n}(f;\Lambda)$ as follows: 
\vspace*{-0.15cm} 
\begin{equation}\label{eqn
} H_{\mathrm{vdm}, n}(f;\Lambda) = H_{\mathrm{coh}, n}(f;\Lambda) + H_{\mathrm{diff}, n}(f;\Lambda),
\end{equation} 
where $H_{\mathrm{coh}, n}(f;\Lambda)$ denotes the spatially coherent component, and $H_{\mathrm{diff}, n}(f;\Lambda)$ denotes the diffuse component. Accordingly, the target dereverberated \ac{VDM} signal is given by
 \vspace*{-0.15cm}
\begin{equation}\label{eqn:target} 
Z_{\mathrm{target}}(f,t) = \sum_{n=1}^{N}H_{\mathrm{coh}, n}(f;\Lambda)  S_{n}(f,t). 
\end{equation}


\section{Proposed Method}
\subsection{DNN Architecture and Training Loss}
\begin{figure}[t]
    \centering
    \includegraphics[width=.89\columnwidth]{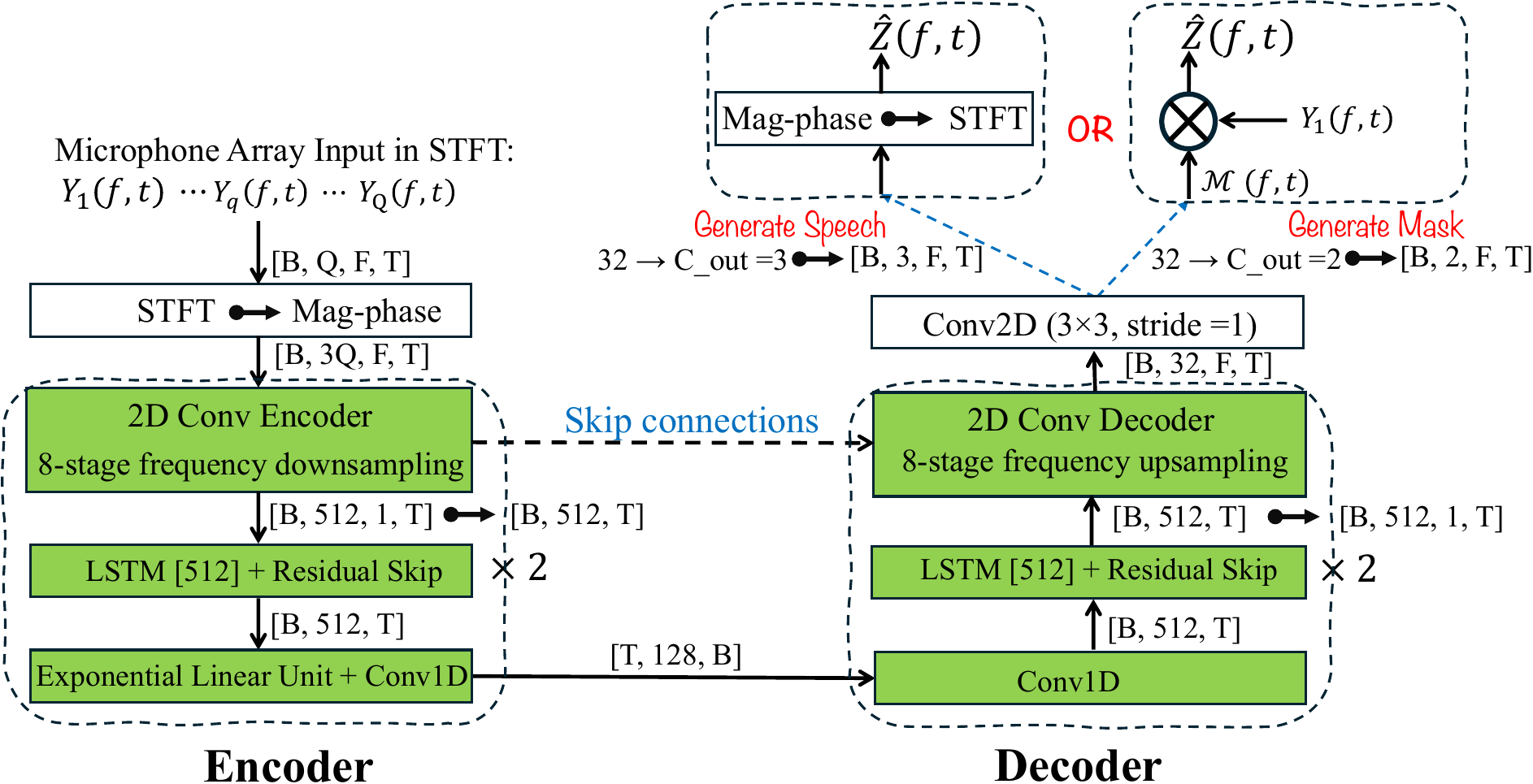}
        \vspace*{-0.1cm} 
    \caption{Generator architecture}
    \label{fig:arch}
    \vspace*{-0.4cm} 
\end{figure}

The GAN-based architecture uses a SEANet-based generator \cite{tagliasacchi2020seanet}, as illustrated in Fig.~\ref{fig:arch}. This design adopts a UNet-like structure in the time-frequency domain, featuring a symmetric encoder–decoder network with skip connections. For the $q$-th microphone, a magnitude-phase representation is computed based on the \ac{STFT} signals as:
 \vspace*{-0.15cm}
\begin{equation} \label{eqn:mag_phase}
    \mathbf{Y}_{q}(f,t) = \Biggl[\,\log |Y_q(f,t)|,\; \frac{\Re(Y_q(f,t))}{|Y_q(f,t)|},\; \frac{\Im(Y_q(f,t))}{ |Y_q(f,t)|}\,\Biggr].
\end{equation}
Concatenating across $Q$ microphones produces an input of size 
$[B, 3Q, F, T]$, where $B$ is the batch size, 
$F$ is the number of frequency bins, and $T$ is the number of time frames. This input is processed by an encoder comprising an initial convolution followed by eight downsampling stages. Each stage includes a residual block~\cite{defossez2022high} and a strided two-dimensional convolution that halves the frequency dimension while maintaining the time dimension. The first four stages incrementally double the channel count, whereas the subsequent four stages retain a constant channel dimension. Upon completion of the final stage, the frequency dimension is reduced to one, yielding a one-dimensional feature sequence. Temporal modeling is performed by a two-layer \ac{LSTM} network with a residual skip connection. The decoder is structured as a mirror of the encoder, employing transposed two-dimensional convolutions for frequency upsampling. At each decoding stage, the corresponding encoder feature map is added element-wise via skip connections, followed by a residual block that refines the combined representation. The final convolution projects the features into a configuration-dependent output space, producing either $[B, 3, F, T]$ for direct dereverberated VDM estimation $\widehat{Z}(f,t)$ in magnitude-phase form or $[B, 2, F, T]$ for complex mask estimation. The complex mask $\mathcal{M}(f,t)$ is then applied to the reference signal $Y_{1}(f,t)$ to obtain the estimated signals, specifically $\widehat{Z}(f,t) = \mathcal{M}(f,t) Y_{1}(f, t)$. The generator that performs dereverberated VDM estimation is referred to as a signal-based UNet, whereas the generator that estimates a complex mask is termed a mask-based UNet.

The loss function of the generator, consistent with \cite{saha2025gan}, is optimized using a weighted combination of four loss terms:
 \vspace*{-0.15cm}
\begin{equation} \label{eqn:total_loss}
    \mathcal{L}_\text{Generator} = \lambda_1\,\mathcal{L}_\text{temp} + \lambda_2\,\mathcal{L}_\text{spec} + \lambda_3\,\mathcal{L}_\text{adv} + \lambda_4\,\mathcal{L}_\text{feat}.
\end{equation}
Here, $\mathcal{L}_\text{temp}$ denotes the $\ell_1$ loss between the target and reconstructed signal waveforms. $\mathcal{L}_\text{spec}$ represents a combination of $\ell_1$ and Frobenius distances computed on Mel and magnitude spectra at multiple resolutions \cite{du2024funcodec}. $\mathcal{L}_\text{adv}$ refers to a hinge-based adversarial loss, while $\mathcal{L}_\text{feat}$ is the $\ell_1$ distance between intermediate feature maps of the discriminator for the target and reconstructed signals. The discriminator architecture utilizes a multi-scale STFT-based network, as described in~\cite{defossez2022high}, with a configuration similar to
\cite{shetu2025leveraging,du2024funcodec }.

\subsection{Training Strategy}

\begin{figure}[t!] 
\centering	
\includegraphics[width=0.599\linewidth]{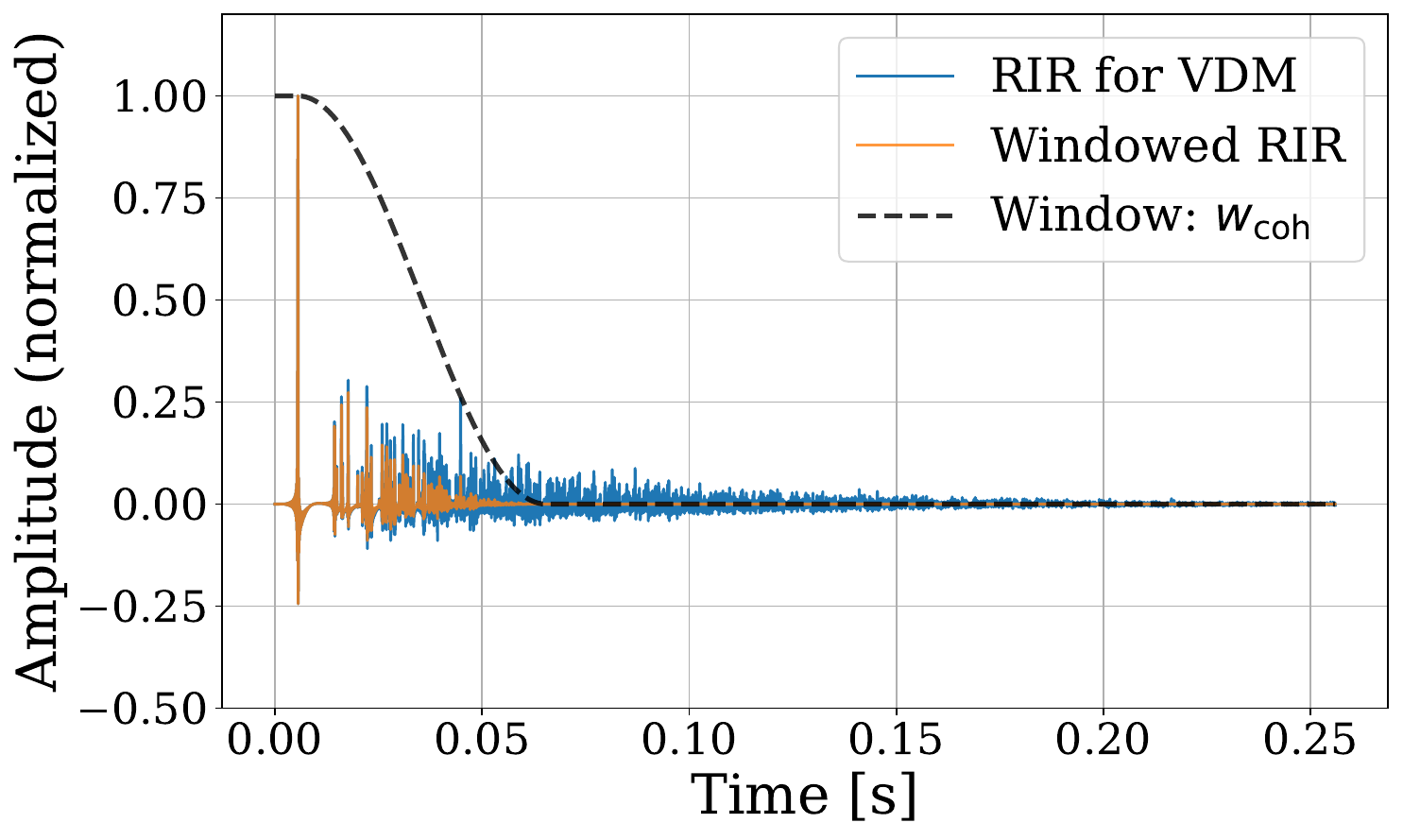}
	\vspace*{-1em}
    \caption{The windowing of the RIR for VDM to preserve the direct sound and early reflections}
	\label{fig: windowing}
    \vspace*{-1.5em}
\end{figure}

In this study, both a $1^{\textrm{st}}$-order Cardioid and a $6^{\textrm{th}}$-order Cardioid are selected as target directivity patterns. A $J^{\textrm{th}}$-order Cardioid directivity pattern \cite{NDF} is adopted as
 \vspace*{-0.15cm}
\begin{equation}\label{eqn:simple_dma_pattern_definition}
\Lambda(\theta, \phi) = (0.5+ 0.5(\sin \phi \sin   \phi_\textrm{s} \cos(\theta - \theta_\textrm{s}) + \cos \phi \cos   \phi_\textrm{s} ) )^{J},
\end{equation} 
where $\theta_\textrm{s}$ and $\phi_\textrm{s}$ specify the target direction of the directivity pattern. The maximum attenuation at the null position of the directivity patterns is set to $-30$~\unit{\decibel} to ensure robust training.

All microphones and sound sources are assumed to lie in the $x$-$y$ plane. To learn the target directivity pattern in a reverberant environment, a random source-array setup with up to three concurrent sources is simulated. The azimuth angle $\theta_{n}$ for the $n$-th speech source relative to the array is randomly selected, and each speech source is assigned a random source-array distance. A room with random dimensions and reverberation time is defined, and the source-array setup is randomly positioned within the room. Based on the positions of the microphones and sources, the corresponding \acp{RIR} \cite{RIRGenerator} are generated, and the microphone signals are computed using \eqref{eqn:mic_sig}. 
 
To compute the training target, the \ac{RIR} for the transfer function $H_{\mathrm{coh}, n}(f;\Lambda)$ is approximated by windowing the corresponding \ac{RIR} for the transfer function $H_{\mathrm{vdm}, n}(f; \Lambda)$, as illustrated in Fig.~\ref{fig: windowing}. The specific definition of the window and corresponding windowing process can be found in \cite{NDF+}.


\section{Experimental Setup}

\subsection{Dataset and Configurations}
\begin{table}
        \setlength\extrarowheight{0.1pt}
        \centering
		\caption{\small{Ranges for reverberant room acoustic settings}}
		\resizebox{.38\textwidth}{!}{
			\begin{tabular}{l c rrr rrr r}
				\toprule
                   \multicolumn{1}{c}{Length} &\multicolumn{1}{c}{Width}&\multicolumn{1}{c}{Height}&\multicolumn{1}{c}{$\textrm{RT}_{60}$}&\multicolumn{1}{c}{Source-array dist.} \\
				\midrule
                     6 - 10~\unit{\metre} & 4 - 8~\unit{\metre}  & 3 - 5~\unit{\metre}  & 0.2 - 0.5~\unit{\s}  & 0.5 - 2.5~\unit{\metre}  \\

				\bottomrule
			\end{tabular}
		}
        \vspace*{-1.5em}
		\label{tab:room_setting}
\end{table}

A four-microphone array was employed, comprising three microphones arranged in a uniform circular array (UCA) with a diameter of 3~cm and one centrally positioned reference microphone. The reference microphone signal was used as the first input channel for the NDDF model. The directivity pattern's target direction ($\theta_s = 0$ and $\phi_s = \frac{\pi}{2}$) was assigned to a selected UCA element, which served as the second input channel for the NDDF model. The array's position within the room was determined using the Monte Carlo Room Impulse Response simulation \cite{MonteCarloRIR}, maintaining a minimum distance of 1.2m from all walls. The source-array distance, room size (length, width, and height), and $\textrm{RT}_{60}$ were randomly sampled from the ranges specified in Table~\ref{tab:room_setting}.

Speech signals for the training and validation sets were obtained from the 'train-clean-360' and 'dev-clean' subsets of the LibriSpeech database \cite{librispeech}, respectively. For the test sets, speech samples were selected from the EARS dataset \cite{richter2024ears}, applying a minimum loudness threshold of $-42$~dBFS \cite{loudness}. All signals were sampled at 16~kHz, and $L=960$ corresponded to a 60~ms duration. Candidate incident angles for the training and validation sets were defined as $\theta_{n} \in \{0^{\circ}, 5^{\circ}, \ldots, 355^{\circ}\}$ and $\theta_{n} \in \{2.5^{\circ}, 7.5^{\circ}, \ldots, 357.5^{\circ}\}$, respectively. The training set consisted of 50,000 samples, and the validation set included 6,000 reverberant samples. Each test set comprised 3,240 samples. Each sample in all datasets lasted 4 seconds. Microphone sensor noise was added at a signal-to-noise ratio (SNR) of 30~dB. The loss weights were set to $\lambda_1 = \lambda_2 = 1$, $\lambda_3 = \frac{1}{9}$, and $\lambda_4 = \frac{100}{9}$, following the original EnCodec configuration~\cite{defossez2022high}.

\subsection{Performance Measures}
\noindent \textbf{Objective metrics}: Since time-domain SDR \cite{vincent2006performance} is unsuitable for generative models without sample-level alignment, we reported frequency-weighted segmental SDR (fwSDR$_{\text{seg}}$), computed in the frequency domain analogously to fwSNR$_{\text{seg}}$ \cite{4389058} but without critical-band energy normalization; the estimation error was treated as distortion. We also computed \ac{PESQ} using the Python \textit{pesqc2} package \cite{pesqc2}, which includes the latest PESQ corrections \cite{torcoli2025navigating}. Both fwSDR$_{\text{seg}}$ and \ac{PESQ} are intrusive metrics requiring target references.

For non-intrusive evaluation, we used \ac{SRMR} \cite{5547575} and $C_{50}$ \cite{kuttruff2016room}. \ac{SRMR} reflects reverberation, while $C_{50}$ measures clarity as the ratio of early ($<50$ ms) to late ($>50$ ms) energy. Here, $C_{50}$ was computed via a DNN-based blind acoustic parameter estimation (BAPE) model \cite{gotz2026multi}.

\noindent $\textbf{Directivity pattern}$: To introduce the proposed dominant bin-based calculation of the directivity pattern, the $n$-th source signal at the reference microphone, $X_{1, n}(f, t)$ in \eqref{eqn:mic_sig}, can be decomposed as
\vspace*{-0.15cm} 
\begin{equation}
X_{1, n}(f, t) = X_{1, n, \textrm{dir}}(f, t) + X_{1, n, \textrm{early}}(f, t)+X_{1, n, \textrm{diff}}(f, t),
\end{equation}
where $X_{1, n, \textrm{dir}}(f, t)$, $X_{1, n, \textrm{early}}(f, t)$, and $X_{1, n, \textrm{diff}}(f, t)$ represent the direct-path, early-reflection, and diffuse components, respectively. Accordingly, $Y_{1, \textrm{dir}}(f, t) = \sum_{n=1}^{N} X_{1, n, \textrm{dir}}(f, t)$, $Y_{1, \textrm{early}}(f, t) = \sum_{n=1}^{N} X_{1, n, \textrm{early}}(f, t)$, and $Y_{1, \textrm{diff}}(f, t) = \sum_{n=1}^{N} X_{1, n, \textrm{diff}}(f, t)$ denote the cumulative direct-path, early-reflection, and diffuse components at the reference microphone. In addition, $H_{\mathrm{coh}, n}(f;\Lambda)$ in \eqref{eqn:target} is decomposed as $H_{\mathrm{coh}, n}(f;\Lambda) = H_{\mathrm{dir}, n}(f;\Lambda) + H_{\mathrm{early}, n}(f;\Lambda)$, where $ H_{\mathrm{dir}, n}(f;\Lambda)$ and $H_{\mathrm{early}, n}(f;\Lambda)$ are the transfer functions corresponding to the direct and early-reflection components, respectively. Let $Z_{\mathrm{early}, n}(f,t) = H_{\mathrm{early}, n}(f;\Lambda) S_{n}(f,t)$ denote the early-reflection components of the target NDDF signal for the $n$-th source. The wideband power ratio $\xi( \theta_n)$ for the $n$-th source is then defined as
\vspace*{-0.15cm} 
\begin{equation}\label{eqn:powerRatio}
\xi( \theta_n)= \frac{ \sum_{f=1}^{F} \sum_{t=1}^{T} \left| \Gamma_{n}(f, t)
 [\widehat{Z}(f,t) -  Z_{\mathrm{early}, n}(f,t)] \right|^2}{\sum_{f=1}^{F} 
 \sum_{t=1}^{T}\left| \Gamma_{n}(f, t) \ Y_{1, \textrm{dir}}(f, t) \right|^2},
\end{equation}
where $\Gamma_{n}(f, t)$ is determined by
\vspace*{-0.15cm} 
\begin{equation}\label{eqn:thereshold}
\Gamma_{n}(f, t) =
\begin{cases}
1, & \text{if } 
\frac{\left| X_{1, n, \textrm{dir}}(f, t) \right|^2}
     {\left| Y_{1, \textrm{coh}}(f, t) \right|^2} \ge \delta
\ \land\ 
\frac{\left| Y_{1, \textrm{coh}}(f, t) \right|^2}
     {\left| Y_{1, \textrm{diff}}(f, t) \right|^2} > 1 \\[0.6em]
0, & \text{else}
\end{cases}
\end{equation}
where $Y_{1, \textrm{coh}}(f, t) = Y_{1, \textrm{dir}}(f, t) + Y_{1, \textrm{early}}(f, t)$, and $\delta \in [0,1)$ is the decision threshold, set to $\delta=0.4$ in this study. Under these conditions, $\Gamma_{n}(f, t)$ determines whether the direct-path of the $n$-th source is dominant in the coherent components, and whether the coherent components surpass the diffuse components for the reference microphone signal. Next, $[\widehat{Z}(f,t) -  Z_{\mathrm{early}, n}(f,t)]$ in \eqref{eqn:powerRatio} is used to approximate the estimated direct-path response of the $n$-th source in the dereverberated \ac{VDM} signal. To obtain the final estimated directivity pattern, the arithmetic mean of $\xi( \theta_n)$ is calculated over all test samples from the same direction $\theta_n$.

\section{Performance Evaluation}

\begin{table*}[t!]
\centering
\caption{Performance comparison of NDDF and corresponding baselines for various $\mathrm{RT}_{60}$. All metrics are positively oriented (higher is better). fwSDR${\text{seg}}$, $C_{50}$, and \ac{SRMR} are reported in dB; \ac{PESQ} is reported on the \ac{MOS} scale. ``Disc." stands for discriminative training. }
\label{tab:combined_tasks}
\sisetup{
    reset-text-series = false,
    text-series-to-math = true,
    mode=text,
    tight-spacing=true,
    round-mode=places,
    round-precision=2,
    table-format=2.2,
    table-number-alignment=center
}
\renewcommand{\arraystretch}{1.2}
\resizebox{0.95\textwidth}{!}{%
\begin{tabular}{l c l *{16}{c}}
\toprule
    Order & Category & Methods
      & \multicolumn{4}{c}{$\mathrm{RT}_{60} = 0.2$~s}
      & \multicolumn{4}{c}{$\mathrm{RT}_{60} = 0.4$~s}
      & \multicolumn{4}{c}{$\mathrm{RT}_{60} = 0.6$~s}
      & \multicolumn{4}{c}{$\mathrm{RT}_{60} = 0.8$~s} \\
    \cmidrule(lr){4-7} \cmidrule(lr){8-11} \cmidrule(lr){12-15} \cmidrule(lr){16-19}
    & & & PESQ & fwSDR\textsubscript{seg} & $C_{50}$ & SRMR
      & PESQ & fwSDR\textsubscript{seg} & $C_{50}$ & SRMR
      & PESQ & fwSDR\textsubscript{seg} & $C_{50}$ & SRMR
      & PESQ & fwSDR\textsubscript{seg} & $C_{50}$ & SRMR \\
    \midrule[1.5pt]
    \multirow{6}{*}{$1^{\text{st}}$}
      & \multirow{2}{*}{Cascaded}
      & DR-SwWPE~\cite{huang2024practical} + DMA~\cite{benesty2018fixed}
      & 2.37 & 18.64 & 30.64 & 5.71
      & 2.24 & 16.84 & 21.47 & 4.71
      & 2.04 & 14.61 & 15.95 & 4.03
      & 1.91 & 12.56 & 11.77 & 3.66 \\
    & & DR-SwWPE~\cite{huang2024practical} + NDF~\cite{NDF}
      & 3.08 & 20.67 & 29.05 & 6.31
      & 2.71 & 19.98 & 18.98 & 5.07
      & 2.36 & 17.97 & 15.00 & 4.31
      & 2.14 & 15.98 & 11.78 & 3.90 \\
    \cmidrule(lr){2-19}
    & \multirow{2}{*}{Disc.}
      & NDDF (FT-JNF~\cite{tesch_insights})
      & \bf{4.34} & \bf{34.61} & \bf{33.22} & 6.47
      & \bf{3.96} & 29.44 & 29.51 & 6.08
      & \bf{3.49} & 26.76 & 27.30 & 5.88
      & \bf{3.07} & 25.34 & 24.66 & 5.70 \\
    & & NDDF (Mask-based UNet)
      & 4.23 & 31.63 & 32.77 & \bf{6.50}
      & 3.86 & 28.32 & 30.45 & \bf{6.17}
      & 3.40 & 25.54 & 28.27 & \bf{6.04}
      & 2.98 & 24.37 & 24.71 & \bf{5.81} \\
    \cmidrule(lr){2-19}
    & \multirow{2}{*}{GAN}
      & NDDF (Mask-based UNet)
      & 4.32 & 34.50 & 32.62 & 6.38
      & 3.86 & \bf{30.00} & 30.68 & 5.81
      & 3.42 & \bf{27.52} & 29.60 & 5.55
      & 3.02 & \bf{25.94} & 27.37 & 5.27 \\
    & & NDDF (Signal-based UNet)
      & 4.31 & 32.72 & 32.67 & 6.37
      & 3.87 & 28.81 & \bf{30.81} & 5.86
      & 3.45 & 26.59 & \bf{29.84} & 5.61
      & 3.06 & 25.37 & \bf{28.10} & 5.38 \\
    \midrule[1.5pt]
    \multirow{5}{*}{$6^{\text{th}}$}
      & Cascaded
      & DR-SwWPE~\cite{huang2024practical} + NDF~\cite{NDF}
      & 2.65 & 16.41 & 29.04 & 6.01
      & 2.37 & 14.71 & 20.11 & 5.10
      & 2.13 & 13.38 & 15.55 & 4.38
      & 1.98 & 12.30 & 12.38 & 4.02 \\
    \cmidrule(lr){2-19}
    & \multirow{2}{*}{Disc.}
      & NDDF (FT-JNF~\cite{tesch_insights})
      & 3.88 & 22.52 & 32.85 & 6.51
      & \bf{3.40} & 18.45 & 29.49 & 6.25
      & 2.96 & 17.15 & 26.08 & 5.94
      & 2.64 & 16.70 & 23.42 & 5.73 \\
    & & NDDF (Mask-based UNet)
      & 3.73 & 20.01 & 33.75 & \bf{6.73}
      & 3.28 & 16.41 & 31.83 & \bf{6.40}
      & 2.89 & 15.11 & 28.09 & \bf{6.03}
      & 2.61 & 14.70 & 24.44 & \bf{5.78} \\
    \cmidrule(lr){2-19}
    & \multirow{2}{*}{GAN}
      & NDDF (Mask-based UNet)
      & \bf{3.89} & 23.44 & \bf{34.13} & 6.39
      & 3.39 & \bf{20.56} & 31.33 & 6.00
      & \bf{3.00} & \bf{19.65} & 28.93 & 5.67
      & \bf{2.72} & \bf{19.07} & 25.83 & 5.42 \\
    & & NDDF (Signal-based UNet)
      & 3.85 & \bf{23.63} & 33.26 & 6.36
      & 3.35 & 20.55 & \bf{31.93} & 6.03
      & 2.97 & 19.49 & \bf{30.26} & 5.73
      & 2.69 & 18.92 & \bf{27.48} & 5.49 \\
    \bottomrule
\end{tabular}%
}
\vspace{-1em}
\end{table*}

\noindent \textbf{Baselines}:
For the $1^{\text{st}}$-order Cardioid target pattern, the baseline was established by cascading a recent real-time multichannel dereverberation algorithm (DR-SwWPE \cite{huang2024practical}) with a $1^{\text{st}}$-order Cardioid \ac{DMA} \cite{benesty2018fixed}, where the DMA was computed using the null-constraint method \cite{benesty2018fixed}. For both $1^{\text{st}}$- and $6^{\text{th}}$-order Cardioid target patterns, another baseline was constructed by cascading DR-SwWPE with the conventional \ac{NDF} \cite{NDF}. For a fair comparison with NDDF, the conventional \ac{NDF} was trained in a simulated reverberant environment using the same settings as those in Table~\ref{tab:room_setting}. To demonstrate the effectiveness of GAN-based training, we trained the mask-based UNet discriminatively to obtain the discriminative NDDF variant. Note that \ac{FT-JNF} was also employed as the DNN architecture for the conventional \ac{NDF}. Accordingly, the NDDF was also implemented using \ac{FT-JNF} with discriminative training.

\begin{figure}[t!]
    \centering
	\begin{minipage}[b]{0.422\linewidth}
		\centering
		\centerline{ \includegraphics[width=\linewidth]{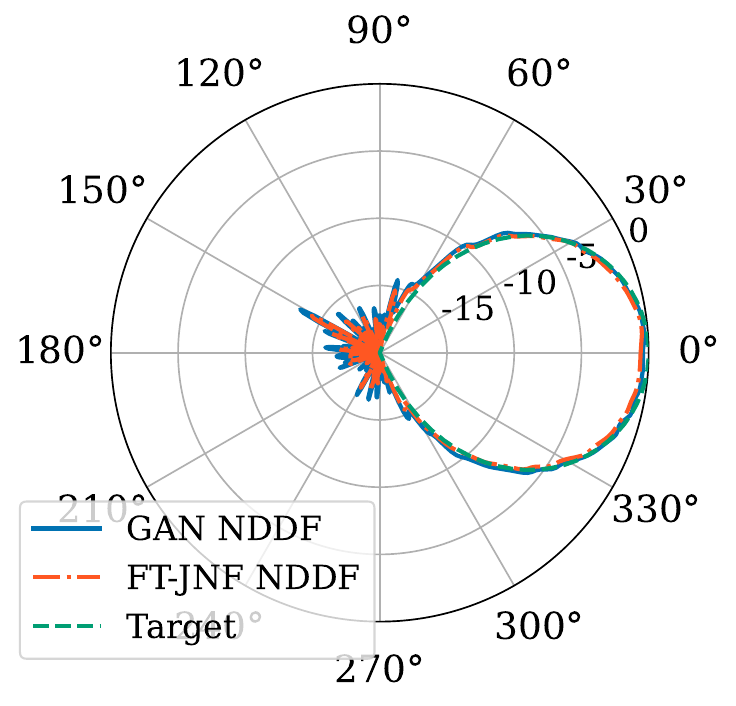}}
		(a)  \footnotesize{$6^{\text{th}}$-order, $\textrm{RT}_{60}= 0.2$~\unit{\s}  }
	\end{minipage}     
    	\begin{minipage}[b]{0.422\linewidth}
		\centering
		\centerline{ \includegraphics[width=\linewidth]{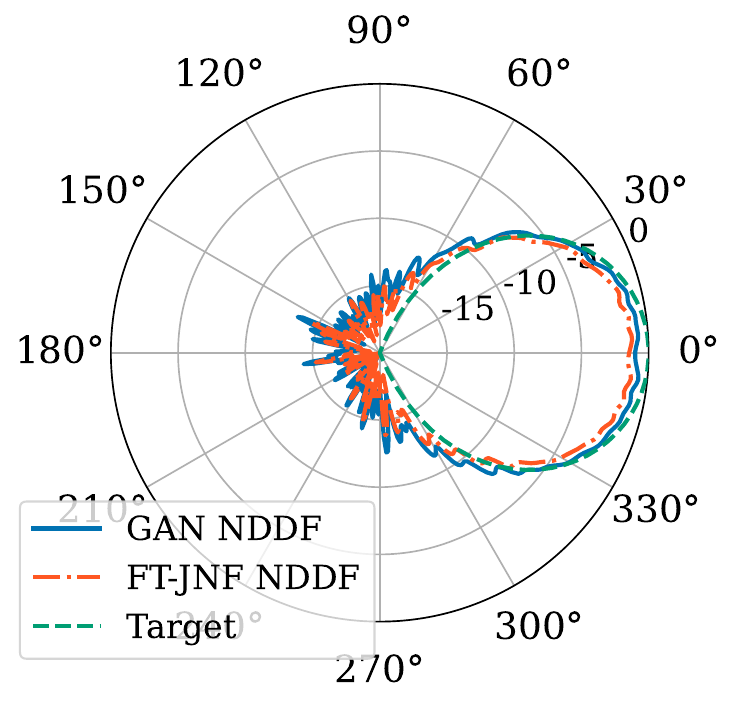}}
		(b)   \footnotesize{$6^{\text{th}}$-order, $\textrm{RT}_{60}= 1.0$~\unit{\s}  }
	\end{minipage}

    
      \caption{Estimated directivity patterns between FT-JNF NDDF (FT-JNF using discriminative training) and GAN NDDF (GAN-based variant with a signal-based UNet). Condition: $6^{\text{th}}$-order target}
	\vspace{-1em}
    \label{fig:wb-bp-rvb}  
\end{figure}


    

\begin{figure}[t!]
    \centering
    	\begin{minipage}[b]{0.45 \linewidth}
		\centering
		\centerline{\includegraphics[width=\linewidth]{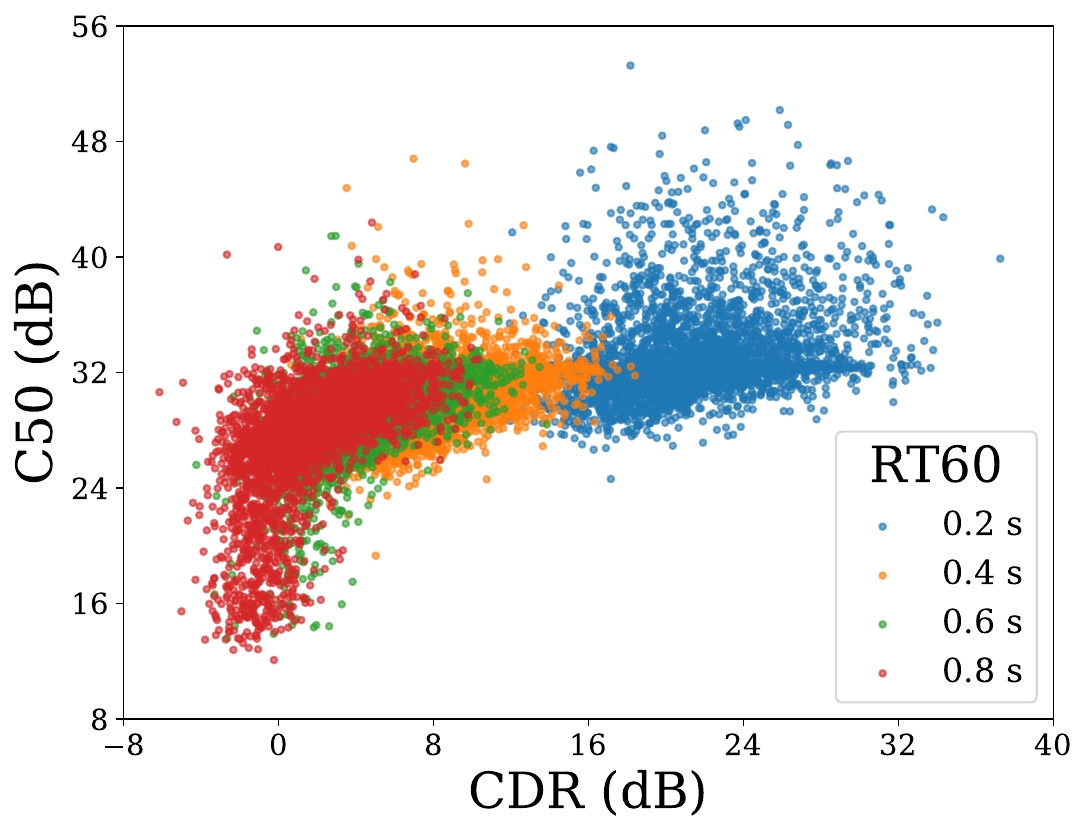}}
		(a) \small{$1^\textrm{st}$-order, FT-JNF~\cite{tesch_insights}   }  
	\end{minipage}
	\begin{minipage}[b]{0.45 \linewidth}
		\centering
		\centerline{\includegraphics[width=\linewidth]{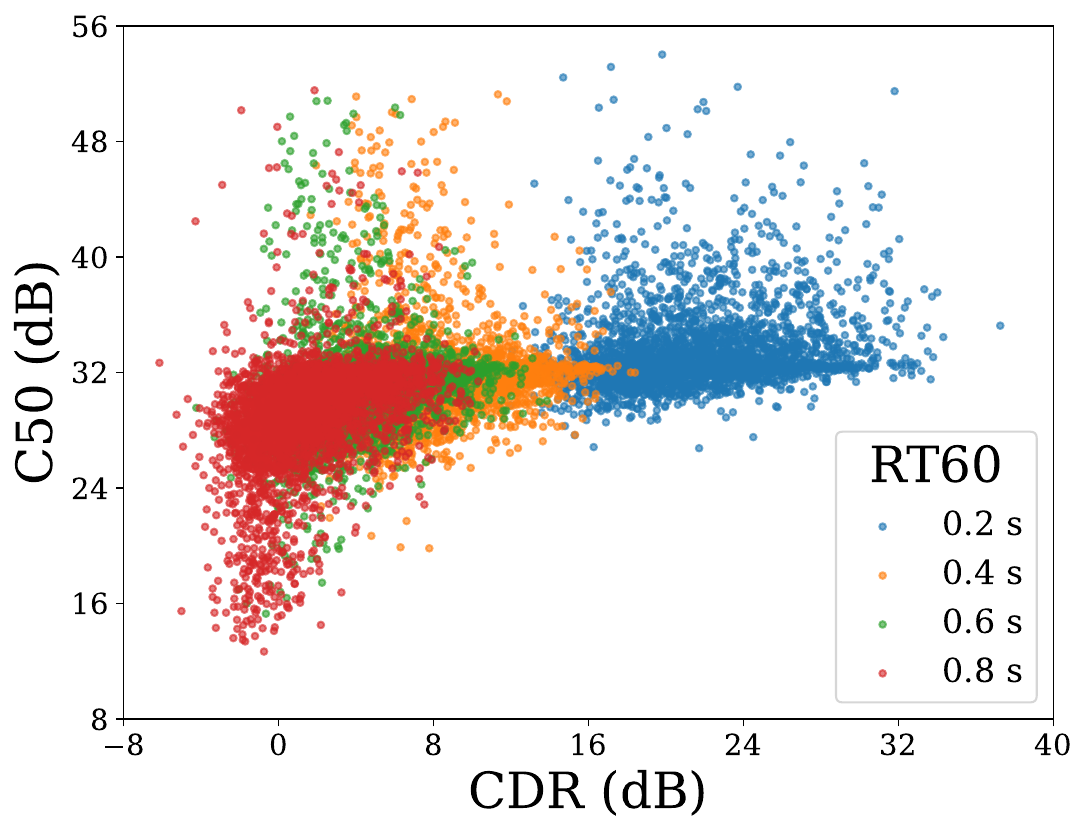}}
		(b) \small{$1^\textrm{st}$-order,  GAN (Signal) }
	\end{minipage}

   	\begin{minipage}[b]{0.45\linewidth}
		\centering
		\centerline{\includegraphics[width=\linewidth]{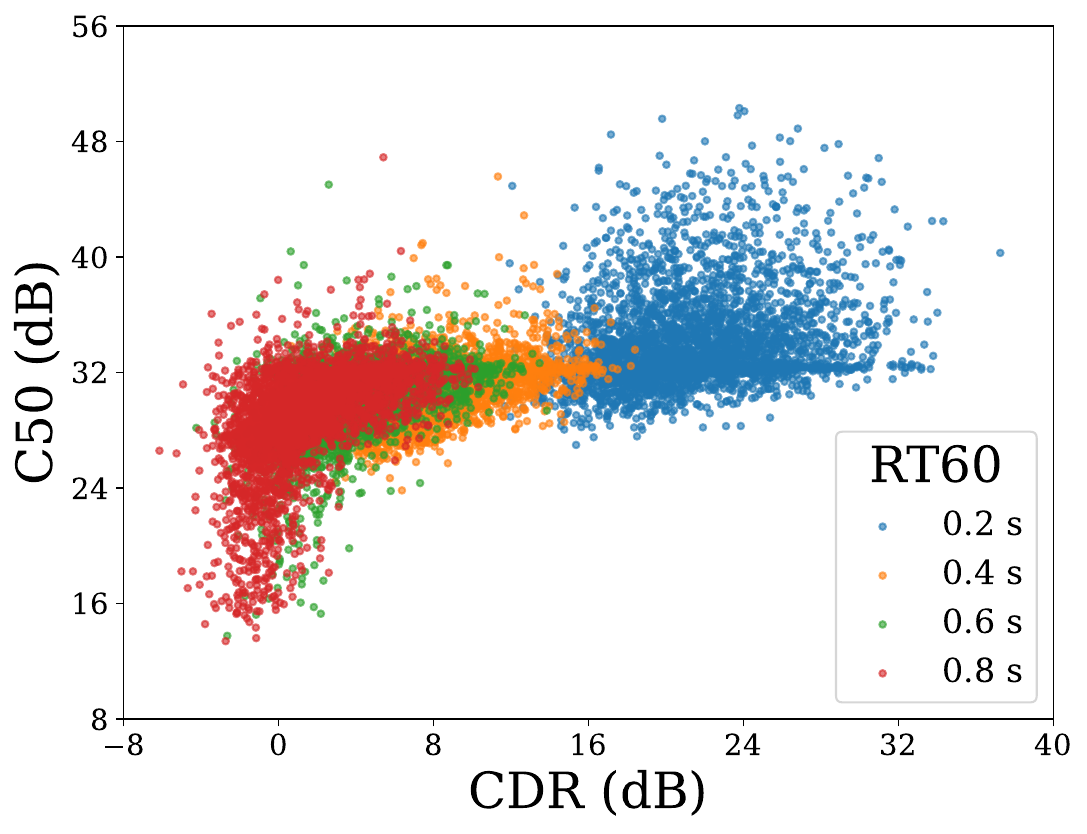}}
		(c) \small{$6^\textrm{th}$-order,   FT-JNF~\cite{tesch_insights}  }
	\end{minipage} 
        \centering
	\begin{minipage}[b]{0.45\linewidth}
		\centering
		\centerline{\includegraphics[width=\linewidth]{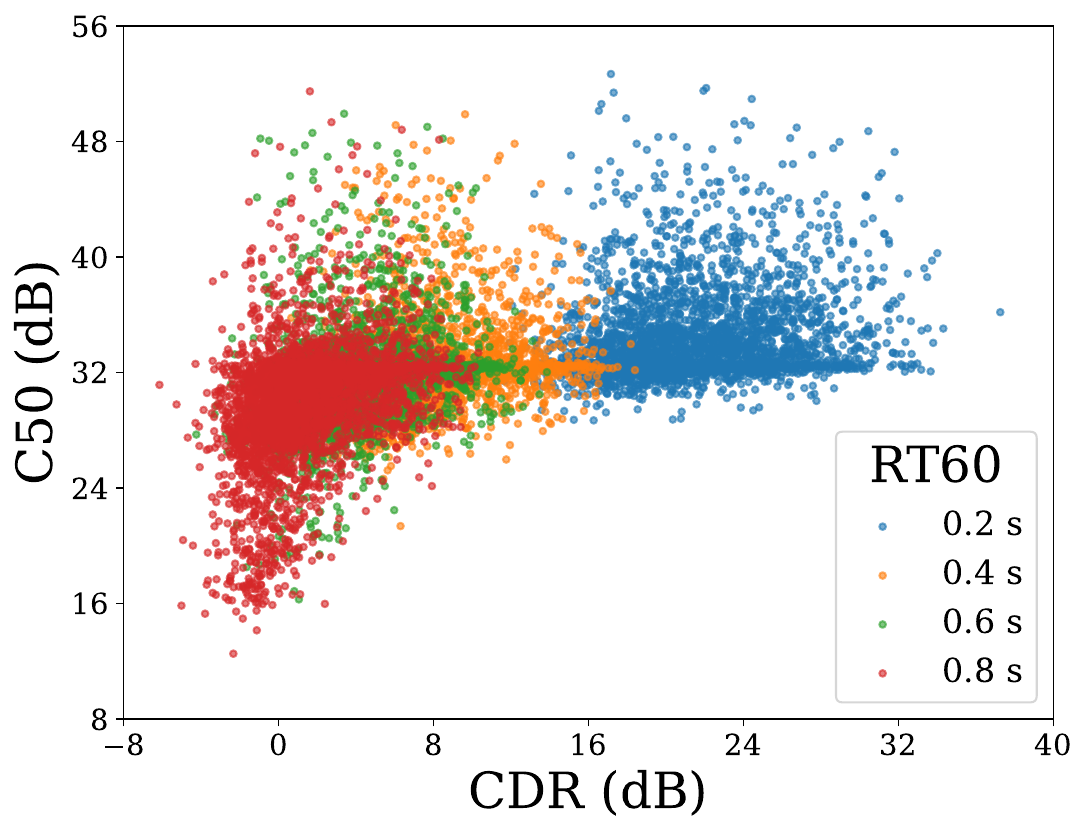}}
		(d) \small{$6^\textrm{th}$-order, GAN (Signal)}
	\end{minipage}
    \vspace{-0.1cm}
\caption{Scatter plots of $C_{50}$ versus CDR for NDDF, evaluated at $\mathrm{RT}_{60} \in \{0.2,\,0.4,\,0.6,\,0.8\}$~s. ``GAN (Signal)'' denotes the GAN-based variant with a signal-based UNet.}
  \label{fig:DRR_SDR_direct}
    \vspace{-1em}
\end{figure}

\noindent \textbf{Overall comparison}: Under varying $\mathrm{RT}_{60}$, we generated test sets with two concurrent sources, with each source randomly sampled from $\theta_n \in \{1.25^\circ, 3.75^\circ, \ldots, 358.75^\circ\}$. Table~\ref{tab:combined_tasks} presents the results for various $\mathrm{RT}{60}$ values. First, NDDF significantly outperforms cascaded methods across all evaluation metrics. Second, for the first-order target, different training paradigms exhibit metric-dependent trade-offs: performance differences on intrusive metrics (i.e., fwSDR\textsubscript{seg} and PESQ) are relatively small, whereas GAN-based training yields clearly higher $C_{50}$, especially under severe reverberation ($\mathrm{RT}_{60}=0.8$~s). In contrast, discriminative training tends to achieve higher SRMR, with the UNet variant attaining the highest SRMR, but performs worse on intrusive metrics, suggesting that high SRMR may result from over-suppression of reverberation.  For the more challenging 6th-order target, we find that GAN-based training methods achieve better overall performance in fwSDR\textsubscript{seg}, PESQ, and $C_{50}$ than discriminative methods, with the advantage becoming more pronounced as reverberation increases. This trend is particularly evident when comparing models with the same mask-based UNet backbone: using this architecture as a GAN generator yields substantially larger gains than training it purely with a discriminative objective. Finally, we observe that the signal-based and mask-based UNets perform comparably for GAN-based training.

\noindent \textbf{Directivity patterns}:
To further examine the differences between discriminative and GAN-based training with respect to directional filtering ability of NDDF, Fig.~\ref{fig:wb-bp-rvb} presents a comparison of the estimated directivity patterns for two representative variants under a sixth-order target: the FT-JNF NDDF (FT-JNF with discriminative training) and the GAN NDDF (GAN-based variant employing a signal-based UNet). Under low-reverberation conditions, both models produce nearly identical patterns. In contrast, under high reverberation, the GAN NDDF more closely approximates the target mainlobe and demonstrates less distortion in the target direction.

\noindent \textbf{Dereverberation analysis}: To specifically assess dereverberation performance, we generated test sets with a single-source setup and placed all sources at the target direction ($\theta_n = 0^\circ$) under varying $\mathrm{RT}_{60}$. For these test sets, we analyzed the $C_{50}$ results of NDDF with respect to the \ac{CDR}, and present the corresponding scatter plots in Fig.~\ref{fig:DRR_SDR_direct}. The \ac{CDR} was measured at the input reference microphone. We observe that $C_{50}$ degrades mildly as the \ac{CDR} decreases, but drops noticeably once the \ac{CDR} falls around $0$~dB; in this low-CDR regime, the GAN-based method exhibits fewer degraded points than FT-JNF. In the CDR range from $0$~dB to $10$~dB, the GAN-based method further yields more samples with $C_{50} > 30$~dB than FT-JNF. As the CDR increases beyond this range, the two methods produce similar scatter distributions. Although the test sets used here differ from those in Table~\ref{tab:combined_tasks}, the observations in this analysis provide a plausible explanation for the $C_{50}$ trends reported therein.

\section{Conclusions}
This paper proposed NDDF to overcome \ac{NDF} limitations in reverberant conditions by reconstructing a dereverberated \ac{VDM} signal. We implemented both discriminative and GAN-based NDDF variants and benchmarked them against established baselines. Across experiments, the GAN-based NDDF achieved the best overall performance. In addition, we introduced an input--output-based directivity pattern estimation method, enabling directivity analysis for signal-mapping spatial filtering approaches.

\bibliographystyle{IEEEbib}
\bibliography{refs}

\end{document}